\documentclass[trackchanges,twocolumn]{aastex7}

\usepackage[T1]{fontenc}
\usepackage{threeparttable}
\usepackage[normalem]{ulem}
\usepackage[version=4]{mhchem} 
\usepackage{placeins}
\usepackage{capt-of}
\usepackage{footnote}
\providecommand{\kms}{km~s$^{-1}\,$}

\usepackage{hyperref}
\begin{document}
% ALMA high resolution water vapor observations reveal the snowline location in the HD 100546 disk

% ho guardato alla lettera, mi sembra un’ottima base! Spingerei un po’ di piu’ nella discussione, conclusioni (che ancora mancano) su:
% questa e’ la prima immagine di main water isotopologue spatially resolved in a transition disk; effectively, it’s the first image of the water snowline, since in this peculiar case the water snowline is pushed out to the cavity wall
% it shows that: water is observable with ALMA in these objects;
% it complements the COMs study with the main oxygen carrier, clearly showing that planets being born in the cavities of these bright TDs will be exposed to chemistry with tons of water vapour
% Questi sono tutti huge points, quindi li dovremo sottolineare per bene.
% Sezione 3: espanderei un attimo quello che si sa su water emission in HD100546

\title{Water vapor emission at the warm cavity wall of the HD 100546 disk as revealed by ALMA}

\author[orcid=0009-0001-2795-8468]{Luna, Rampinelli}
\affiliation{Dipartimento di Fisica, Universit\`a degli Studi di Milano, Via Celoria 16, 20133 Milano, Italy}
\email[show]{luna.rampinelli@unimi.it} 
\author[orcid=0000-0003-4689-2684]{Stefano, Facchini}
\affiliation{Dipartimento di Fisica, Universit\`a degli Studi di Milano, Via Celoria 16, 20133 Milano, Italy}
\email{stefano.facchini@unimi.it}
\author[orcid=0000-0003-3674-7512]{Margot, Leemker}
\affiliation{Dipartimento di Fisica, Universit\`a degli Studi di Milano, Via Celoria 16, 20133 Milano, Italy}
\email{margot.leemker@unimi.it}
\author[orcid=0000-0001-8061-2207]{Andrea, Isella}
\affiliation{Department of Physics and Astronomy, Rice University, 6100 Main Street, MS-108, Houston, TX 77005, USA}
\affiliation{Rice Space Institute, Rice University, 6100 Main Street, MS-108, Houston, TX 77005, USA}
\email{ai14@rice.edu}
\author[orcid=0000-0003-2045-2154]{Pietro, Curone}
\affiliation{Departamento de Astronomía, Universidad de Chile, Camino El Observatorio 1515, Las Condes, Santiago, Chile}
\email{pcurone@das.uchile.cl}
\author[orcid=0000-0002-7695-7605]{Myriam, Benisty}
\affiliation{Max-Planck Institute for Astronomy (MPIA), Königstuhl 17, 69117 Heidelberg, Germany}
\email{benisty@mpia.de}
\author[orcid=0000-0001-9549-6421]{Elizabeth, Humphreys}
\affiliation{Joint ALMA Observatory (JAO),Alonso de Córdova 3107, Vitacura 763-0355, Casilla 19001, Santiago, Chile}
\affiliation{European Southern Observatory (ESO),Alonso de Córdova 3107, Vitacura 763-0355, Santiago, Chile}
\email{ehumphre@eso.org}
\author[0000-0003-1859-3070]{Leonardo, Testi}
\affiliation{Alma Mater Studiorum Università di Bologna, Dipartimento di Fisica e Astronomia (DIFA), Via Gobetti 93/2, 40129 Bologna, Italy}
\email{leonardo.testi@unibo.it}

%\author[orcid=0000-0000-0000-0001,sname='North America']{Tundra North America}
%\altaffiliation{Kitt Peak National Observatory}
%\affiliation{University of Saskatchewan}
%\email[show]{fakeemail1@google.com}  

%\author[orcid=gname=Bosque, sname='Sur America']{Forrest Sur Am\'{e}rica} 
%\altaffiliation{Las Campanas Observatory}
%\affiliation{Universidad de Chile, Department of Astronomy}
%\email{fakeemail2@google.com}

%% Use the \collaboration command to identify collaborations. This command
%% takes an optional argument that is either a number or the word "all"
%% which tells the compiler how many of the authors above the command to
%% show. For example "\collaboration[all]{(DELVE Collaboration)}" wil include
%% all the authors above this command.
%%
%% Mark off the abstract in the ``abstract'' environment. 
\begin{abstract}
We present spatially resolved ALMA observations of the water line at 183~GHz in the disk around the Herbig star HD~100546. The water vapor emission peaks at the inner edge of the warm dust cavity, located $\sim15$ au from the central star. We attribute this to thermal desorption at the water snowline, shifted outward at the dust cavity wall directly heated by the intense radiation. This represents the first spatially resolved image of the water snowline using ALMA observations of the main water isotopologue in a protoplanetary disk. The water emission morphology peaking inside the first dust ring is consistent with previous ALMA detections of oxygen-bearing complex organic molecules in the disk, including thermally desorbed methanol. %This reveals a chemically rich, oxygen-dominated gas-phase environment inside the inner dust ring, influencing the atmospheric composition of forming planets.
These findings signal that warm cavities of transition disks provide ideal targets to directly reconstruct the spatial distribution of water vapor and the snowline location with ALMA, and directly connect water vapor emission to ice desorption of complex organic species. \end{abstract}

%% Keywords should appear after the \end{abstract} command. 
%% The AAS Journals now uses Unified Astronomy Thesaurus (UAT) concepts:
%% https://astrothesaurus.org
%% You will be asked to selected these concepts during the submission process
%% but this old "keyword" functionality is maintained in case authors want
%% to include these concepts in their preprints.
%%
%% You can use the \uat command to link your UAT concepts back its source.

%% From the front matter, we move on to the body of the paper.
%% Sections are demarcated by \section and \subsection, respectively.
%% Observe the use of the LaTeX \label
%% command after the \subsection to give a symbolic KEY to the
%% subsection for cross-referencing in a \ref command.
%% You can use LaTeX's \ref and \label commands to keep track of
%% cross-references to sections, equations, tables, and figures.
%% That way, if you change the order of any elements, LaTeX will
%% automatically renumber them.

\section{Introduction}\label{sec:intro}

Water is of primary interest in astronomy, as it plays a fundamental role in the emergence of life on Earth, liquid water being an efficient solvent that facilitates the development of chemical complexity. Understanding how water was delivered to Earth is a central question in astronomy \citep{morbidelli2012}. However, answering this question is challenging, as it is closely intertwined with the complex chemical and physical history of the building blocks of planets and their interplay with water in the natal environments, the protoplanetary disks \citep{vandishoeck2021}.

Water observations in the protoplanetary disk stage are extremely valuable, as they offer a unique opportunity to observationally characterize the link between the atmospheric composition of the evolved exoplanetary population and their birthplace \citep{cridland2016}. In addition, the water in these sources is likely formed before the central star did, thus injecting all systems with a similar water abundance at the time of their formation \citep{tobin2023,andreu2023,leemker2025d2o}. Being a primary oxygen carrier, water plays a crucial role in determining the chemical make-up of the planet-forming material, by often dominating the C/O ratio in the gas and ice phases (\citealt{oberg2011}; \citeyear{oberg2023}; \citealt{eistrup2018}). Beyond the water snowline, at a midplane temperature below $\sim 150$~K, the water freezes on dust grains, enriching the ice with oxygen sequestered from the gas phase. The location of the snowline not only regulates the chemical composition of the disk reservoir but also promotes planet formation by facilitating grain growth and coagulation \citep{Schoonenberg2017,drazkowska2023}. 

\begin{figure*}[ht!]
    \centering
    \includegraphics[scale=0.27]{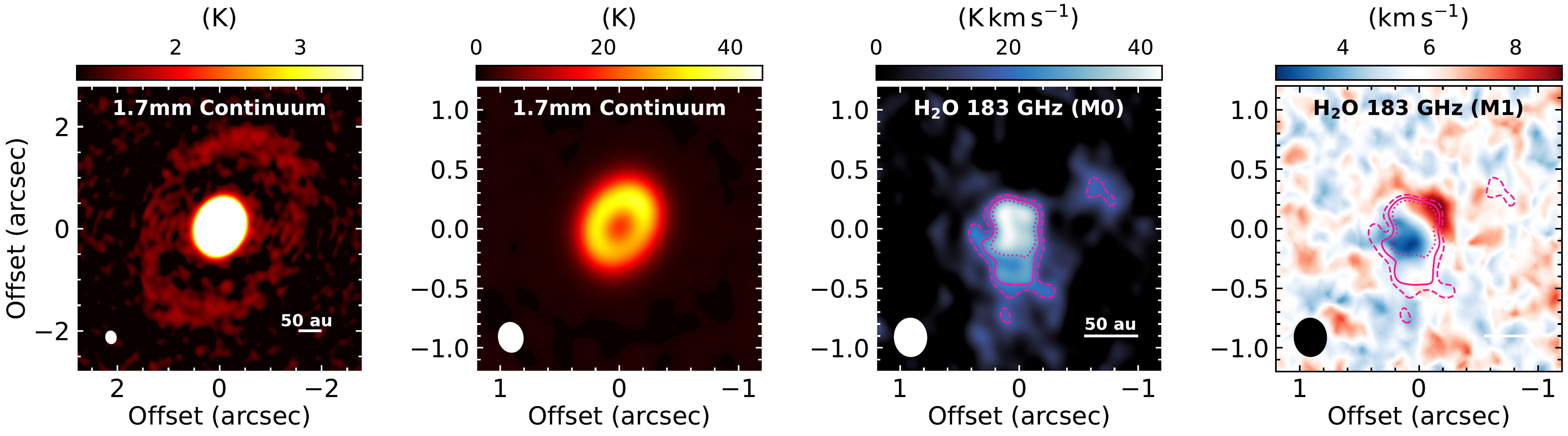}
    \caption{First and second panels: 1.7 mm continuum maps of the HD~100546 disk showing the faint outer ring (first), and a zoom in on the inner ring (second). Third panel: integrated intensity map of the \ce{H2O} 183~GHz line emission. The intensity is expressed as velocity integrated brightness temperature. Pink lines show 3, 4, and 5 $\sigma$ contours, where $\sigma$ is evaluated as standard deviation inside an annulus between $2\farcs3$ and $4\farcs0$. Fourth panel: intensity weighted velocity map of the 183~GHz water line. The color scale is centered at the systemic velocity of 5.7~\kms (white color). The ellipse in the bottom left of each panel represents the beam.}
    \label{fig:M0}
\end{figure*}

Water observations in disks have been performed mainly with infrared campaigns, and have confirmed the main water reservoirs predicted by disk thermo-chemical models \citep[see, e.g.,][and references therein]{pontoppidan2010, banzatti2017, vandishoeck2021}. Water gas is expected in three main regions in protoplanetary disks: (1) inside the water snowline, typically within a few astronomical units from the central star. This is a warm reservoir of thermally desorbed water at temperatures above $\sim 150$~K, with high abundances typically $\gtrsim10^{-4}$ \citep{woitke2009, vandishoeck2014}; (2) in the outer disk. This is the cold component ($T \lesssim 100$~K) photodesorbed from icy grains, at intermediate layers where UV photons can penetrate. The abundance can reach up to $10^{-6}$ and it is regulated by photodissociation and water self-shielding \citep{bergin2010, hogerheijde2011}; (3) in the disk surface layer. This is the hot component reformed in the gas phase at $T\gtrsim300$~K. The high temperature required to unlock the gas-phase reaction pushes this region to higher layers, where the abundance can reach up to $10^{-4}$ as a result of the balance between photodissociation and self-shielding \citep{bethell2009}. We refer the reader to \cite{vandishoeck2021,pirovano2022} for a detailed description.

%Infrared observations of protoplanetary disks revealed the before-mentioned water reservoirs with distinct abundances and excitation properties, suggesting different underlying chemical and physical conditions and production pathways \citep[see, e.g., ][]{pirovano2022,temmink2024,romero-mirza2024, banzatti2025}.
A first characterization of the water emission in protoplanetary disks was provided by IR campaigns with Spitzer, revealing correlations with stellar and disk properties, such as the stellar temperature and luminosity, accretion luminosity, disk dust mass and size, but also dynamical processes such as the inward drift of icy pebbles  \citep[see e.g.][]{pontoppidan2010,salyk2011,najita2013,banzatti2020}. Higher resolution water spectra of the inner disk of protoplanetary disks obtained with the JWST revealed multiple temperature contributions associated to different water reservoirs \citep{banzatti2023b, kamp2023}, with the component consistent with the water sublimation temperature becoming more prominent at longer wavelengths \citep[$\sim23~\mu\mathrm{m}$,][]{banzatti2023, temmink2024}. The cold water reservoir photodesorbed in the outer disk was detected by Hershel in the disks of TW~Hya and HD~100546, revealing a weaker emission than previously predicted by models, as a result of the settling of icy grains over time \citep{bergin2010}. 

However, IR observations with current facilities lack the spatial resolution to resolve the water emission and directly constrain the snowline location, which is typically within only a few astronomical units for T~Tauri stars and within $\sim10$~au for Herbig stars \citep{woitke2009,notsu2017}. \cite{tobin2023} showed the first direct constraint of the water snowline location (at $\sim80$~au) from ALMA observations of heavier water isotopologues in the disk around the outbursting V883~Ori star. \cite{facchini2024} showed the first spatially and spectrally resolved ALMA observations of the main water vapor isotopologue in the protoplanetary disk of HL~Tau. These results demonstrate the potential of ALMA in revealing the content, distribution, and location of the snowline of water vapor in disks. However, even though less important than in the IR, dust absorption can occult mm-wave water line emission arising from the disk midplane. Further imaging of water transitions at mm wavelengths in protoplanetary disks with optically thin inner regions can circumvent this problem \citep{leemker2025HLTau}.

%Moreover, this is highlighting the importance of developing a strong synergy between space- and ground-based telescopes to characterize water vapor where planetary systems are built. In this context, new breakthrough water observations informed by models predictions (\citealt{vandishoeck2014,notsu2016}, \citeyear{notsu2017}, \citeyear{notsu2018}; \citealt{kim2024,hasegawa2024}) can reveal new strategies and push the observational frontiers to directly unravel the fundamental role of water in planet formation. 

In this work we present the first spatially resolved ALMA observations of water vapor in a disk hosting a wide dust cavity. %, directly revealing the snowline location at the warm cavity wall of the disk around HD~100546. 
The source is described in Sect.~\ref{sec:source}, while observations used to perform the analysis, self-calibration and image reconstruction are outlined in Sect.~\ref{sec:obs}. We discuss the main results in Sect~\ref{sec:discussion}, and summarize the main conclusions in Sect.~\ref{sec:conclusions}.

\section{The source}\label{sec:source}
HD~100546 is a Herbig star (M$_\star =  2.4$~M$_\odot$, \citealt{walsh2017}, L$_\star =$ 23.5~L$_\odot$, \citealt{vioque2018}), at a distance of 110~pc (\citealt{gaia2016}, \citeyear{gaia2023}). Two dust rings at $\sim25$ and $\sim200$~au have been detected in ALMA continuum observations \citep{walsh2014, pineda2019, fedele2021}, while scattered light observations revealed small- and large-scale spiral arm structures \citep{garufi2016,follette2017,sissa2018}. Both wide orbit and close-in forming planet candidates have been suggested from sub-millimeter and infrared observations, asymmetric SO emission, and CO kinematic signatures \citep{currie2015,sissa2018,casassus2019,pineda2019,brittain2019,perez2020,booth2023}.

HD~100546 shows a line-rich spectrum with a variety of molecules detected with ALMA, among which complex organic molecules (COMs) such as methanol and methyl formate, showing ring shape emission and/or azimuthal variations \citep{booth2023,keyte2023,Booth2024HD100546,leemker2024,evans2025}. In particular, methanol has an excitation temperature of $\sim152$~K inside 110~au, which is consistent with thermal desorption origin \citep{evans2025}. Thermally-desorbed water is therefore expected to be detected in the same region as methanol, as they have similar sublimation temperatures \citep{wakelam2017,minissale2022}. \cite{leemker2024} modeled ALMA line emission observations of the HD~100546 disk with DALI \citep{bruderer2012dali1, bruderer2013dali2}, predicting the water snowline location to be at $\sim15$~au, at the inner edge of the first dust ring.

Gas-phase water was previously detected in the disk around HD~100546 with Herschel \citep{vandishoeck2021,pirovano2022}. Both cold and warm lines with upper state energies ranging from $\sim50$ to $\sim1000$~K were targeted, but only the two cold lines at 557~GHz (E$_\mathrm{u}=61$~K) and 1113 GHz (E$_\mathrm{u}=53$~K) were detected, and attributed to the cold photodesorbed water reservoir. The emitting region was inferred to be outside 40~au from the spectrally resolved line profiles \citep[]{vandishoeck2021}. 

\cite{pirovano2022} modeled the \textit{Herschel} observations of HD 100546 using DALI \citep{bruderer2012dali1, bruderer2013dali2} and constrained the water abundance in the three reservoirs discussed in Sect.~\ref{sec:intro}.  Their fiducial model predicts low water abundances in the thermally desorbed component ($<10^{-9}$) and in the hot reservoir formed in the gas phase ($<10^{-10}$), while the photodesorbed component at intermediate layers outside the dust ring at $\sim40$~au needs to be relatively abundant ($\sim3\times10^{-9}$) to reproduce the observations.
%\cite{vandishoeck2021} highlighted that the non-detection of warmer water lines, which are expected to dominate the emission inside $\sim40$~au, could be a result of the dust ring being optically thick at \textit{Herschel} wavelengths, thus hindering the detection of most of the water emission. Moreover, the low ($< 10^{-9}$) water abundance inside the inner dust cavity inferred by \cite{pirovano2022} is consistent with the location of the water snowline at the outer edge of the gas cavity at $\sim15$~au, as predicted by \cite{leemker2024} from thermochemical modeling of the HD~100546 disk. This is also supported by the centrally peaked \ce{HCO^+} emission in ALMA band~7 observations \citep{Booth2024HD100546}, which do not allow to spatially resolve the inner $\sim30$~au.

Finally, \cite{honda2016} detected an absorption feature at 3$\mu$m with Gemini in the HD~100546 disk, associated with water ice in the surface layer in the outer gap between 40 and 120~au, which is in line with the results presented by \cite{pirovano2022}. 

\section{Observations}\label{sec:obs}

In this Letter, we analyzed ALMA band~5 observations of HD~100546 from project 2023.1.01431.S (PI S.~Facchini), which covers the \ce{H2O} $3_{1,3}-2_{2,0}$ line at $\nu = 183.310$~GHz.
The observations consist of three Execution Blocks (EBs), obtained in C-5 configuration on the 2024-06-30 (EB0 and EB1) and 2024-07-02 (EB2), with a maximum baseline of 2.5~km. The time on source for each EB was 47 minutes, for a total of 2.35 hours in exquisite weather conditions corresponding to a Precipitable Water Vapor (PWV) ranging from 0.21 to 0.45~mm across the EBs. The spectral setup is made of two spectral windows (spw) centered on the water $3_{1,3}-2_{2,0}$ line at $\nu = 183.310$~GHz and the \ce{H^13CO^+} $2-1$ line at $\nu = 173.507$~GHz, respectively, and two continuum spws centered at 172.0 and 185.5~GHz with a 1.875~GHz bandwidth each. Each of the two line spws has a bandwidth of 117~MHz, and a native spectral resolution of 122~kHz ($\sim 0.2$~\kms) for the \ce{H2O} spw and 30.5~kHz (0.05~\kms) for the \ce{H^13CO^+} spw. After the ALMA Pipeline cross-calibration, we self-calibrated the data following the Pipeline designed for the exoALMA Large Program \citep{loomis2025}, and using the \texttt{CASA} software version 6.5.4 \citep{CASA2022}. %We describe the steps we followed to self-calibrate the data below.
\begin{figure*}[ht!]
    \centering
    \includegraphics[scale=0.6]{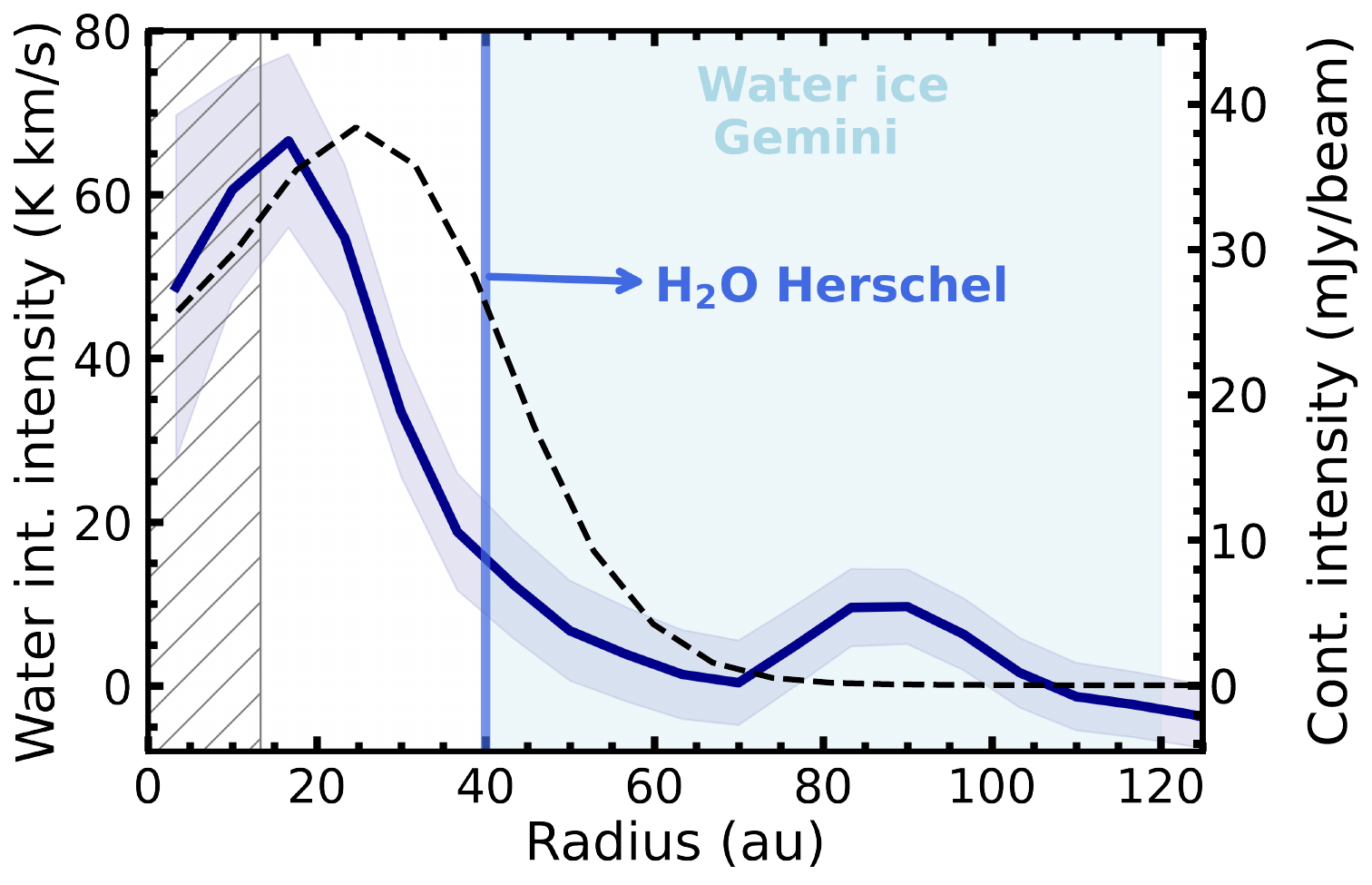}
    \caption{Radial profile of the integrated intensity of the water emission (blue line) compared to the continuum one (dashed black line). Blue ribbons show the uncertainty on the water radial profile. The light blue region indicates the radial extent of the 3$\mu$m water ice absorption band from Gemini observations \citep{honda2016}, and the vertical blue line shows the inner edge of the water emission inferred from \textit{Herschel} observations \citep[]{vandishoeck2021,pirovano2022}. The hatched region corresponds to the beam semi-major axis of the water cube used to extract the profile. The negative trend of the radial profile beyond 100~au seems to be an artifact due to negative sidelobes in the PSF.}
    \label{fig:radial_profile}
\end{figure*}

\subsection{Self-calibration} \label{subsec:selfcal}

After flagging lines within a $\pm 15$~\kms range around the systemic velocity, we averaged the data into 125~MHz wide channels. We applied a first round of phase-only self-calibration to individual EBs, combining spws and scans on EB-long time intervals. We then performed a $uv$-plane alignment as described by \cite{loomis2025}. The flux rescaling was not performed at this stage of the self-calibration since phase decoherence was found in the data (see \citealt{loomis2025} for more details). We then concatenated all EBs and applied self-calibration iteratively on progressively shorter time intervals (EB-long, 360s, 120s, 60s, 18s, and 6s). Only spws and polarizations (no scans) were combined for time intervals shorter than 360s. Phase self-calibration was performed using models produced through \texttt{tclean}, cleaning down to 6$\sigma$. An elliptical mask was used during the cleaning, with Position Angle $\mathrm{PA}=146.0^\circ$, a major axis of $2\farcs5$ to include the faint outer ring at $\sim 200$~au, and a minor axis of $2\farcs5 \times \cos{(i)}$, where $i = 41.7^\circ$ is the disk inclination \citep{Booth2024HD100546}.

Phase self-calibration improved phase coherence and flux rescaling was not necessary since flux offsets among different EBs were within 4\%. We therefore proceeded with two rounds of amplitude self-calibration, using EB-long and scan-long intervals, respectively, and combining spws and polarizations. For this step we cleaned down to $1\sigma$. We then applied the gain solutions and phase shifts (in the original order) to the full dataset, including spectral lines. Finally, we binned the data into 30s intervals and performed the continuum subtraction using the \texttt{uvcontsub} task.
We report a continuum flux of $290$~mJy inside the de-projected circular mask used during the self-calibration, a noise RMS of 21.5~$\mu$Jy~beam$^{-1}$, and a peak S/N ratio of 2240 at the end of the self-calibration. 

Particular care was taken to estimate the noise RMS in the water spw, because of the low transmission caused by the atmospheric telluric line of water. The spectrum of the RMS and its impact on the water spectrum is discussed in detail in Appendix~\ref{app:noise}.

\subsection{Imaging}\label{subsec:imaging}
The analyses presented in this work were applied to spectral cubes of the water line reconstructed through the \texttt{CASA} software \citep{CASA2022} using the \texttt{tclean} task \citep{hogbom1974aperture, cornwell2008}. We used the \texttt{multiscale} deconvolver with scales [0,5,10,20,30] pixels, a pixel size 
of $0\farcs01$, and a flux threshold of $3\sigma$. To place the CLEAN components we used the same elliptical mask built for the self-calibration and used to image the continuum. 

To find the best balance between sensitivity and spatial resolution, we imaged the line with \texttt{Briggs} weighting spanning a range of robust parameters between $0.5$ (beam size = $0\farcs25 \times 0\farcs20$, RMS $\sim$ 8~mJy~beam$^{-1}$) and $2.0$ (beam size = $0\farcs32 \times 0\farcs27$, RMS $\sim$ 5~mJy~beam$^{-1}$). However, note that the RMS varies substantially in the spectral range of the water emission, see Appendix~\ref{app:noise} for more details. Similarly, to find the best balance between spectral resolution and S/N, we imaged the cube with a channel width of $0.6$~\kms and $1.2$~\kms. For the results presented in this Letter we used cubes both with a channel width of $0.6$ and $1.2$~\kms, and imaged with a robust parameter of 0.5 and 2.0. Figure~\ref{fig:channel_maps} in Appendix~\ref{app:channel_maps} shows the channel maps of the water line spectral cube, imaged with natural weighting and a channel width of 1.2~\kms. We applied a correction for the so-called JvM effect \citep{jorsater1995} following \cite{czekala2021}, to correctly recover the flux of the water line.

\subsection{Moment maps, radial profile and spectrum}\label{subsec:spec+prof+m0}
\begin{figure*}[ht!]
    \centering
    \includegraphics[scale=0.35]{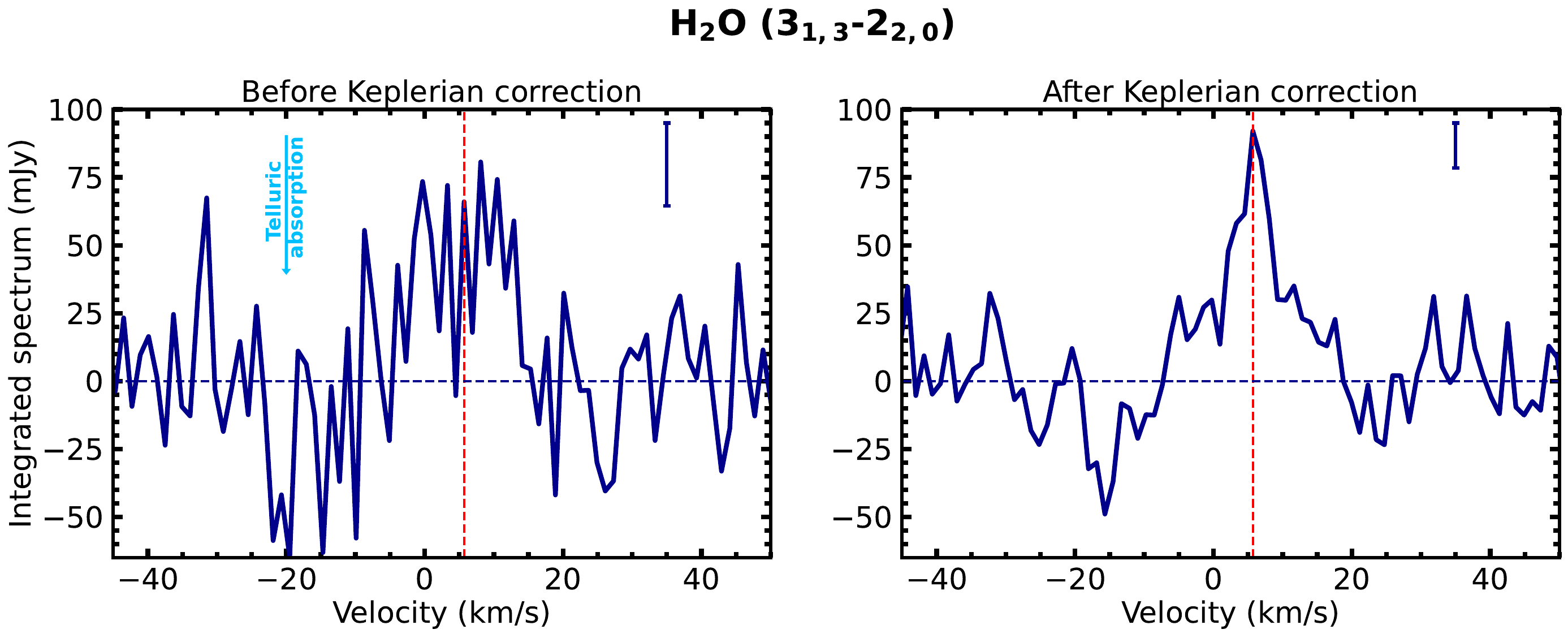}
    \caption{Disk-integrated spectrum of the \ce{H2O} line, extracted from a de-projected circle with a radius of 100~au (left panel), and after correcting for the Keplerian rotation (right panel). The vertical red dashed line indicates the systemic velocity of $5.7$~\kms, while the blue line on the top right side of each panel is the uncertainty. The dip in the spectrum in the left panel at $\sim-21$~\kms is due to the low transmission at the peak of the atmospheric telluric water line (see Appendix~\ref{app:noise} for details on the bandpass calibration and its effect on the retrieved water spectra).}
    \label{fig:spectrum}
\end{figure*}

We calculated the integrated intensity map showed in the middle panel of Fig.~\ref{fig:M0}, from the cube imaged with natural weighting to maximize the sensitivity. We integrated the cube between $-6.3$~\kms and $13.5$~\kms, after visual inspection of the data with no additional masking, using the \texttt{bettermoments} package \citep{teague2019bettermoments}. The water integrated intensity (moment 0) map is compared to the continuum one in Fig.~\ref{fig:M0}. Most of the water emission originates from within the bright continuum ring at $\sim 15$~au, while there is a fainter azimuthally asymmetric component emitting from outside the continuum ring at $\sim 85$~au towards south/west. The right panel of Fig.~\ref{fig:M0} shows the intensity weighted velocity map, obtained by integrating in the same velocity range used to calculate the moment 0 map. The (moment 1) map shows the typical blue- and red-shifted pattern around the systemic velocity of 5.7~\kms, expected for a disk in Keplerian rotation.

We calculated the water line flux to be $469 \pm 165$~mJy~\kms by spatially integrating the velocity integrated intensity map inside an elliptical mask with a major axis of 100~au and a minor axis of $100\times\cos{(i)}$, using the disk geometrical parameters (PA = $146.0^\circ$ and $i=41.7^\circ$) from \cite{Booth2024HD100546}. The uncertainty is evaluated as the standard deviation of the flux measured in the same elliptical mask used to evaluate the line flux, but outside the emitting area of the integrated intensity map. We took the maximum number of ellipses that was possible to place on the image without overlap, performing the procedure on images that are not primary beam corrected, to ensure uniform noise. The flux was extracted from the JvM-corrected cube to avoid overestimating the line flux, while the uncertainty was evaluated from the non-JvM corrected image to avoid underestimating the noise \citep{rampinelli2024}. 

From the integrated intensity map of the water emission we extracted the radial profile of the integrated intensity, shown in Fig.~\ref{fig:radial_profile}, using the \texttt{GoFish} package \citep{teague2019gofish}. The uncertainty on the radial profile was extracted as:
\begin{equation}
    \sigma_\mathrm{radial\,profile} = \frac{\sigma_\mathrm{chan} \cdot\Delta v \cdot N_\mathrm{{chan}}}{\sqrt{N_\mathrm{{beam/bin}}}},
\end{equation}
where $\sigma_\mathrm{chan}$ is the standard deviation over the first five and last five channels of the cube used to extract the moment 0 map, $\Delta v$ is the channel width, $N_\mathrm{{chan}}$ is the number of channels over which we integrated to extract the moment 0, and $N_\mathrm{{beam/bin}}$ is the number of beams in each annulus over which we calculated the radial profile of the integrated intensity (which is set to one if $N_\mathrm{{beam/bin}} < 1$).

We show the radial profile obtained from the cube imaged with a robust parameter of 0.5 in favor of a higher spatial resolution to highlight possible radial features. The inner peak of the water emission coincides with the inner edge of the dust cavity wall, as visible from the comparison of the solid blue and dashed black profiles in Fig.~\ref{fig:radial_profile}. The outer azimuthally asymmetric emission observed in the integrated intensity map in Fig.~\ref{fig:M0} is also visible in the radial profile in Fig.~\ref{fig:radial_profile}, which shows an outer emission bump peaking at $\sim 85$~au. We will further discuss the emission morphology of water vapor in Sect.~\ref{sec:discussion}.

We extracted the disk-integrated spectrum from the cube imaged with natural weighting, using the \texttt{GoFish} package \citep{teague2019gofish}. The water line spectrum obtained by integrating over elliptical mask used to extract the water flux is shown in the left panel of Fig.~\ref{fig:spectrum}. The spectrum shows the typical double-horn feature expected for a Keplerian disk, and centered at the systemic velocity of $5.7$~\kms, indicated by the vertical dashed red line. The S/N associated to the peak of the spectrum is 2.6. The uncertainty is 30 mJy and it is evaluated as standard deviation of the 88 signal-free velocity bins (blue errorbar in the top right corner of Fig.~\ref{fig:spectrum}). We then extracted the spectrum from the same region but correcting for the Keplerian rotation of the disk \citep{teague2019gofish}, as shown in the right panel of Fig.~\ref{fig:spectrum}. The spectral shifting boosts the peak S/N to 5.7, with the uncertainty decreasing to 16.5 mJy. 

We highlight that the width of the spectrum in the right panel of Fig.~\ref{fig:spectrum} should not be interpreted as the water linewidth, since it is affected both by beam and spectral smearing due to the limited spatial (beam major axis $\sim$ 35~au) and spectral (channel width = 1.2~\kms) resolution of the cube used to extract it. Moreover, additional sources of uncertainty possibly affecting the Keplerian shift are associated to the assumed geometrical parameters (inclination, PA, emitting surface), stellar mass, and distance of the source.

\section{Discussion}\label{sec:discussion}

The presented observations show the first spatially resolved detection of the main water vapor isotopologue with ALMA in a transition disk. %This result highlights the potential of ALMA of revealing the water vapor content and distribution where planets form and build their atmospheres. 
As anticipated in Sect.~\ref{subsec:spec+prof+m0}, the water emission at 183.31~GHz shows an inner bright peak at $15$~au. In the following sections we discuss this result in light of the previous dust and line observations of the disk around HD~100546.

\subsection{Water vapor at the warm cavity wall}\label{subsec:results_cavity}
The peak of the radial profile of the water integrated intensity in Fig.~\ref{fig:radial_profile} is at the cavity wall of the dust disk, at a radius of ($15\pm5$)~au. The astrometric accuracy on the peak of the profile in Fig.~\ref{fig:radial_profile} is estimated as $\theta_\mathrm{FWHM} / \mathrm{S/N} / 0.9$\footnote{See Sect. 10.5.2 \textit{Astrometric Observations \& Astrometric Accuracy} in the \href{https://almascience.nrao.edu/proposing/technical-handbook/}{ALMA Cycle 12 Technical Handbook}}, which implies a $2\sigma$ offset between the disk cavity wall at $\sim15$~au and the peak of the dust ring at $\sim25$~au. We suggest this emission originates from thermally desorbed water at the snowline.

The location of the water snowline is typically only a few astronomical units away from the central star for T~Tauri stars and within $\sim10$ au also for warmer Herbig disks, as the high water binding energy implies a high sublimation temperature of $\sim150$~K. However, HD~100546 is a transition disk around a Herbig star, with an intrinsically warmer temperature structure \citep{leemker2022} and a central dust and gas cavity \citep{pineda2019,perez2020,wolfer2023,Booth2024HD100546}. These physical conditions result in an efficiently irradiated cavity wall reaching temperatures higher than the water sublimation temperature \citep{leemker2024}, and pushing the water snowline to larger radii. This is also consistent with the thermo-chemical model from \cite{leemker2024} predicting the water snowline to be located at $\sim15$~au at the cavity wall. 

While less severe than at IR wavelengths \citep{leemker2025HLTau}, optically thick dust can partially hide water vapor emission. The rapid drop outside $\sim20$~au in the radial profile of the water integrated intensity in Fig.~\ref{fig:radial_profile} could also be a consequence of the optically thick dust ring at $\sim25$~au, instead of water freeze-out. However, high spatial resolution ALMA band~7 continuum observations show that the peak of the continuum brightness temperature (at $\sim25$~au) is $\sim80$~K, well below the water sublimation temperature of $\sim150$~K, suggesting that the water snowline is located at the inner edge of the dust ring at $\sim15$~au. This is also supported by radiative transfer models showing that the midplane dust temperature steeply drops below $\sim 150$~K at the cavity wall \citep{pineda2014,pineda2019,keyte2023,leemker2024}.

The thermal desorption origin of the observed water emission at the dust cavity wall is supported by the fact that multiple oxygen-rich molecules including methanol have been suggested to thermally sublimate in the HD~100546 disk \citep{Booth2024HD100546, evans2025}. %In particular, \cite{evans2025} showed that methanol (which has a similar desorption temperature to water) is thermally desorbed in the inner disk inside $\sim110$~au, with an excitation temperature $T_\mathrm{ex} = 152$~K.

To summarize, the evidence outlined above points to the conclusion that the water snowline is located at the cavity edge of the dust disk at $\sim15$~au in the HD~100546 disk.

\subsection{Comparison with \textit{Herschel} observations}\label{subsec:herschel}
Previous observations of water vapor in this source with \textit{Herschel} only revealed cold water transitions ($E_\mathrm{up} = 53$ and $61$~K), associated to the photodesorption layer outside $\sim40$~au. 
Hot ($E_\mathrm{up} > 300$~K) and warm ($E_\mathrm{up} \sim 200-300$~K) water lines were also targeted with \textit{Herschel} but not detected in the disk of HD~100546 \citep{vandishoeck2021,pirovano2022}.

The hot water lines with high $E_\mathrm{up} \,(\gtrsim 300$~K) are expected to trace the hot water reservoir produced in the gas phase in the upper layers and inside the dust cavity (see Sect.~\ref{sec:intro}), where photodissociation limits the water abundance. Water self-shielding has been shown to mitigate this effect when the water column density is high ($\gtrsim10^{18}$~cm$^{-2}$, \citealt{bethell2009,bosman2022,calahan2022}). However, HD~100546 is a transition disk with a large dust cavity which is also depleted in gas. The right panel of Fig~\ref{fig:model_water_ncr} in Appendix~\ref{app:model_ncr} represents the radial profile of the water column density from the thermochemical model by \cite{leemker2024}, which shows that the water column density in the disk cavity is expected to be low enough to allow efficient photodissociation. We highlight that this model includes photodissociation only in the vertical direction, and not in the radial one. The non-detection of the hot \textit{Herschel} lines with $E_\mathrm{up} = 432, 552, \mathrm{and}\, 843$~K which trace the hot water reservoir in the upper layers and inside the cavity can be thus explained by the efficient photodissociation \citep{pirovano2022}.

On the other hand, the water column density is higher at the disk cavity edge at $\sim15$~au where the water snowline is expected to be located (see the right panel of Fig.~\ref{fig:model_water_ncr}), thus favoring water self-shielding against photodissociation. The warm \textit{Herschel} lines with $E_\mathrm{up}= 194, 300, \mathrm{and}\, 323$~K and the ALMA line with $E_\mathrm{up} = 205$~K, which are expected to trace closer to the water snowline, are therefore expected to be detected.

These warm \textit{Herschel} lines could go undetected if they are sub-thermally excited, as their critical density is higher than the local gas density. The left panel of Fig.~\ref{fig:model_water_ncr} shows the critical density of the \textit{Herschel} lines presented by \cite{pirovano2022} (solid lines) and of the ALMA 183~GHz line (dashed line), as a function of temperature, calculated following \cite{faure2024}, under the optically thin assumption, and neglecting absorption, stimulated emission, and background emission (see Appendix~\ref{app:model_ncr}). The critical densities of the warm \textit{Herschel} lines with $E_\mathrm{up}= 194, 300, \mathrm{and}\, 323$~K are $\sim10^9$~cm$^{-3}$ at $\sim150$~K, and approximately a factor of three higher than the ALMA line. However, we highlight that these warm \textit{Herschel} lines have high Einstein coefficients $\log_{10}(A_\mathrm{ul}[\mathrm{s}^{-1}]) = -0.59, -0.48, \mathrm{and}\, -0.61$, and are therefore expected to be optically thicker than the ALMA 183~GHz line with $\log_{10}(A_\mathrm{ul}[\mathrm{s}^{-1}]) = -5.44$. In the optically thick regime, photon trapping can consistently lower the critical density \citep{shirley2015}. This, together with the fact that the gas density is expected to be $>10^9$~cm$^{-3}$ at the disk cavity wall \citep{leemker2024}, suggests that sub-thermal excitation cannot explain the non-detection of the warm \textit{Herschel} lines against the detection of the ALMA line.  

The non-detection of warm water vapor with \textit{Herschel} could be due to the high continuum optical depth of the inner dust ring at \textit{Herschel} wavelengths with respect to ALMA, potentially hiding water emission in this region in the FIR. On the other hand, the optical depth of dust is much lower at ALMA wavelengths thus allowing to detect the warm 183~GHz ALMA line ($T \gtrsim 150$~K) emitting at the dust cavity edge.

Finally, we report a marginal detection (peak $\mathrm{S/N} = 4$ in the channel maps) of azimuthally asymmetric water emission peaking at $\sim85$~au in the West and South sides of the disk (see the 8.10 \kms channel in Fig.~\ref{fig:channel_maps}). In this context, we note that HD~100546 has a dynamically active disk with large-scale spiral arms \citep{garufi2016}. The faint bump is also visible in the radial profile of the integrated intensity (Fig.~\ref{fig:radial_profile}) but the S/N is too low ($\sim2$) to robustly conclude on its significance. Interestingly, photodesorption of water ice has been proposed in the cold outer gap between the two dust rings at $\sim25$ and $\sim200$~au \citep{honda2016,pirovano2022}.

\subsection{Water vapor emission and the molecular complexity of HD 100546}\label{subsec:other_mol}

\cite{Booth2024HD100546} presented ALMA line emission observations of a rich sample of molecular tracers in the disk of HD~100546, most of them showing a double ringed emission similar to what is observed in ALMA continuum observations. 
\cite{leemker2024} showed that chemistry rather than the underlying density structure is responsible for the observed molecular rings. 
%In this context, \cite{leemker2024} performed thermochemical modeling of the HD~100546 disk with DALI, and showed that the location of the molecular rings cannot be reproduced by a model with a single gas gap depth, since CO isotopologues, HCN, and \ce{HCO^+} suggest a shallow gas gap, while CN and \ce{C2H} a deep gas gap. 
In particular, oxygen and carbon rich molecules show the peak of the emission just inside or outside the inner dust ring at $\sim25$~au, respectively. In addition, most of the molecules show a second peak just outside the outer dust ring at $\sim200$~au \citep{Booth2024HD100546}. This morphology suggests a radially varying C/O ratio, higher at the dust rings, where \ce{C2H} and CN show bright emission \citep{leemker2024}. This picture is supported by the detection of methanol and SO inside the inner dust cavity \citep{evans2025}, and it is also consistent with the water morphology presented in this work. As shown in Fig.~\ref{fig:radial_profile}, the inner bright peak of water emission at $\sim15$~au is just inside the inner dust ring, while the outer faint water emission peaking at $\sim85$~au is within the gas gap, both anti-correlating with the dust rings where the C/O ratio is proposed to be above one \citep{leemker2024}.

The observed water emission morphology is expected to anti-correlate with the \ce{HCO+} emission, the latter being a major destruction pathway for water \citep{phillips1992,bergin1998,leemker2021}. \ce{HCO+} shows a centrally peaked emission and an outer ring co-located with the outer dust ring \citep{Booth2024HD100546}. The observed faint water emission at $\sim85$~au is therefore consistent with a region of fainter \ce{HCO^+} emission. While the radial profile of the water integrated intensity in Fig.~\ref{fig:radial_profile} shows marginal evidence of a decreasing trend inside $\sim15$~au, the spatial resolution of both the \ce{H2O} and \ce{HCO^+} observations does not allow to robustly conclude about their anti-correlation inside the inner dust cavity. The marginal inner drop in \ce{H2O} could also be due to a drop in the gas density, as the HD~100546 shows an inner gas cavity. On the other hand, \ce{HCO^+} is a good tracer of the ionization rate and if it became optically thicker than water in the inner gas cavity, it would follow the drop in the gas density \citep{leemker2024}.
However, we highlight that the detection of water vapor is consistent with the low \ce{HCO^+}/CO column density ratio extracted for the HD~100546 disk \citep{Booth2024HD100546}, which could either result from a low ionization rate or the presence of gas-phase water.

\section{Conclusions}\label{sec:conclusions}

In this work we present the first spatially resolved observations of the main water vapor isotopologue with ALMA in a transition disk.
Our observations of the HD~100546 disk show:
\begin{enumerate}
    \item water emission peaking at the warm cavity wall of the transition disk. The high temperature in the dust cavity induced by the intense radiation field pushes the snowline to larger radii, allowing water to thermally desorb at the inner edge of the dust cavity;
    \item the first \textit{direct} image of the water snowline in a protoplanetary disk from spatially resolved observations of the main water isotopologue;
    \item the gas phase distribution of water vapor, complementing previous observations of O-rich COMs in the transition disk. This reveals that planets forming inside the warm cavity of the transition disk are exposed to a water-rich chemistry which can leave a crucial imprint on the atmospheric composition of planets forming in the disk;
    \item the potential of ALMA of tracing the water vapor distribution in planet-forming disks thanks to the extremely favorable weather conditions reached at the observatory site, and the high resolution and sensitivity that can be achieved. 
\end{enumerate}

%% Please use the acknowledgment and contribution environments. This will 
%% be anonomyized when the "anonymous" style option is used. 
\begin{acknowledgments}
We thank the referee for the thorough review and helpful comments.

This paper makes use of the following ALMA data: 

ADS/JAO.ALMA 2023.1.01431.S.

ALMA is a partnership of ESO (representing its member states), NSF (USA) and NINS (Japan), together with NRC (Canada), MOST and ASIAA (Taiwan), and KASI (Republic of Korea), in cooperation with the Republic of Chile. The Joint ALMA Observatory is operated by ESO, AUI/NRAO and NAOJ.

L.R., S.F., and M.L. are funded by the European Union (ERC, UNVEIL, 101076613). Views and opinions expressed are however those of the authors only and do not necessarily reflect those of the European Union or the European Research Council. Neither the European Union nor the granting authority can be held responsible for them. S.F. also acknowledges financial contribution from PRIN-MUR 2022YP5ACE.
A.I. acknowledge support from the National  Aeronautics and Space Administration under grant No. 80NSSC18K0828.
P.C. acknowledges support by the ANID BASAL project FB210003.
M.B. has received funding from the European Research Council (ERC) under the European Union’s Horizon 2020 research and innovation programme (PROTOPLANETS, grant agreement No. 101002188).
LT acknowledges financial support from the European Research Council via the ERC Synergy Grant “ECOGAL” (project ID 855130).

\end{acknowledgments}

%\begin{contribution}
%%This section gives authors the space to recognize author contributions. The text inside this environment is NOT counted towards the total word quanta. At a minimum, manuscripts are expected to include this text:

%All authors contributed equally to the Terra Mater collaboration.

%% But authors are expected to provide more specific details, e.g. 
%%
%%SC was responsible for writing and submitting the manuscript.
%%WWM came up with the initial research concept and edited the manuscript.
%%OTS obtained the funding and edited the manuscript.
%%EBF provided the formal analysis and validation. He also edited the manuscript.
%%GEH Supervised the undergraduates, wrote the software and administers the project github and Zenodo repositories.
%%
%% Authors can use the Contributor Role Taxonomy (CRediT) at
%% https://credit.niso.org
%% for ideas on how write a good statement tailored to their needs.

%\end{contribution}

%% To help institutions obtain information on the effectiveness of their 
%% telescopes the AAS Journals has created a group of keywords for telescope 
%% facilities.
%
%% Following the acknowledgments section, use the following syntax and the
%% \facility{} or \facilities{} macros to list the keywords of facilities used 
%% in the research for the paper.  Each keyword is check against the master 
%% list during copy editing.  Individual instruments can be provided in 
%% parentheses, after the keyword, but they are not verified.
\facilities{ALMA}

%% Similar to \facility{}, there is the optional \software command to allow 
%% authors a place to specify which programs were used during the creation of 
%% the manuscript. Authors should list each code and include either a
%% citation or url to the code inside ()s when available.
\software{\texttt{CASA} v6.5.4 \citep{CASA2022},  
          \texttt{numpy} \citep{harris2020}, 
          \texttt{matplotlib} \citep{hunter2007}, \texttt{bettermoments} \citep{teague2019bettermoments}, \texttt{gofish} \citep{teague2019gofish}
          }

%% Appendix material should be preceded with a single \appendix command.
%% There should be a \section command for each appendix. Mark appendix
%% subsections with the same markup you use in the main body of the paper.
%%
%% Each Appendix (indicated with \section) will be lettered A, B, C, etc.
%% The equation counter will reset when it encounters the \appendix
%% command and will number appendix equations (A1), (A2), etc. The
%% Figure and Table counter will not reset.

\appendix

\section{Noise in the water spw} \label{app:noise}
\begin{figure*}[ht!]
    \centering
    \includegraphics[scale=0.33]{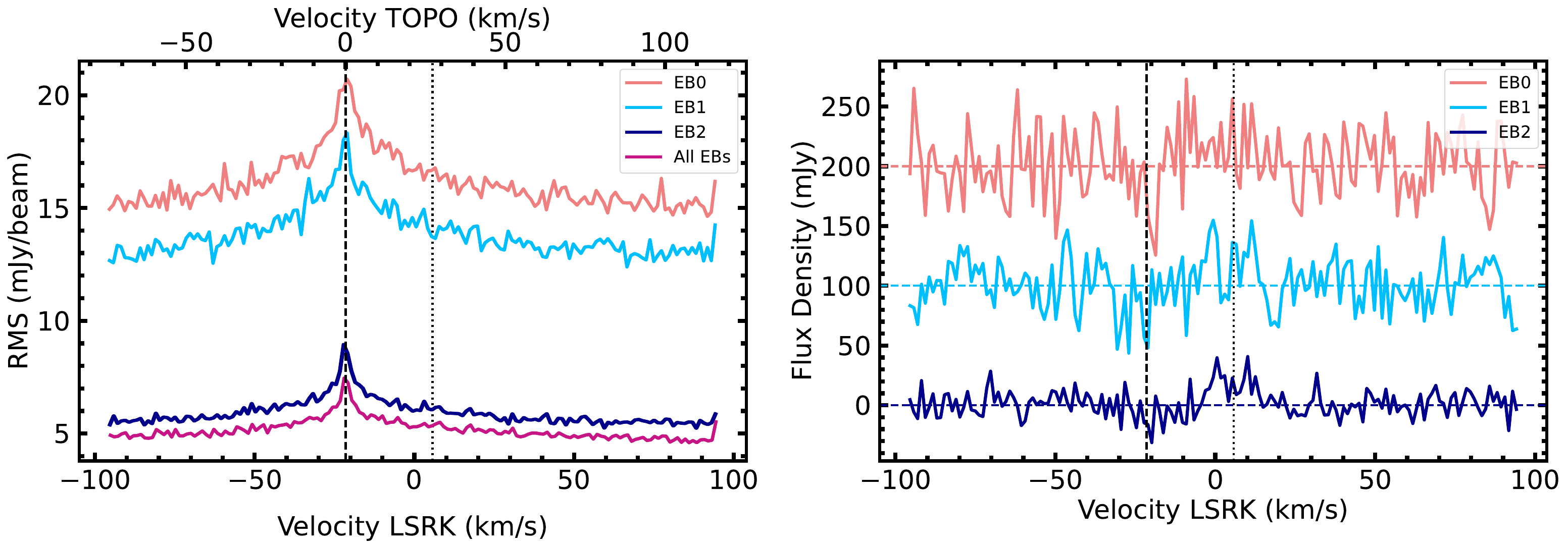}
    \caption{Left: RMS as a function of velocity extracted from a circle with a 500~au radius, for the three EBs and combining all EBs. Right: Water spectrum extracted from a circle with a 50~au radius, for the three EBs (EB0 and EB1 have been shifted by 100 and 200\,mJy for clarity). Non-JvM corrected cubes imaged with natural weighting and a channel width of 1.2~\kms were used. The vertical dashed lines correspond to the systemic velocity in the LSRK frame and the vertical dotted lines to $v = 0$~\kms in the TOPO frame.}
    \label{fig:rms}
\end{figure*}

We report a spectral dependence of the noise RMS in the water spw. The left panel of Fig.~\ref{fig:rms} shows the noise RMS evaluated inside a circle with a radius of 500~au, as a function of velocity. The noise RMS was extracted from non-JvM corrected cubes (to avoid underestimating the noise) imaged with natural weighting and a channel with of 1.2~\kms.  The vertical dotted line shows the systemic velocity of 5.7~\kms (in LSRK velocity frame), while the vertical dashed line corresponds to a velocity of 0~\kms in the topocentric frame. The x-axes show the velocity in LSRK (bottom) and topocentric frame (top). The strong spectral dependence of the noise level is due to the lower transmission at the peak of the water telluric line, which is shifted with respect to the LSRK systemic velocity of the science target, peaking at $\sim-21$~\kms in LSRK frame. As expected, the peak of the noise rms is correctly centered at $0$~\kms in the topocentric frame, with the Lorentzian wings clearly recognizable throughout the spectral window. The observed velocity shift in the LSRK frame with respect to the topocentric frame at the location of ALMA is consistent with the observing dates and times, and with the source coordinates. In particular, the three EBs were taken on 2024-06-30 at 20:55 UTC, 2024-06-30 at 22:24 UTC, and 2024-07-02 at 21:07 UTC. The small difference in dates only induces a small shift in LSRK velocity.

EB0 and EB1 were taken in worse weather conditions than EB2, with the former showing a mean PWV of 0.4~mm, and the latter 0.2~mm. The noise RMS is significantly higher for EB0 and EB1 than for EB2 (see the left panel of Fig.~\ref{fig:rms}), due to the exponential dependence with the atmospheric transmission within the telluric line. Due to the lower signal-to-noise ratio during the first two EBs, the bandpass calibration of the water spw on EB0 and EB1 was performed averaging over 10 and 6 velocity channels, respectively, while no averaging was applied to EB2. This has a direct consequence in the retrieved continuum-subtracted spectrum of the water spw. The right panel of Fig.~\ref{fig:rms} shows the spectrum of the water line, for the three EBs separately (EB0 and EB1 were shifted up for clarity), inside a circle with a 50~au radius. The spectra show a small dip (also visible in Fig.~\ref{fig:spectrum}) at the peak of the spw rms. As expected from the different weather conditions and spectral averaging during bandpass calibration for EB0 and EB1, the spectrum from EB2 shows the smallest dip. 

We highlight that this effect could be particularly important when scheduling water observations with ALMA (in particular for the 183~GHz water line), depending on the observing date and science target, since the trough in the atmospheric transmission could partially overlap with the water line of the astrophysical source of interest, thus resulting in both an elevated rms level and artificial absorption features.

\section{Water channel maps} \label{app:channel_maps}

\begin{figure*}[ht!]
    \centering
    \includegraphics[scale=0.24]{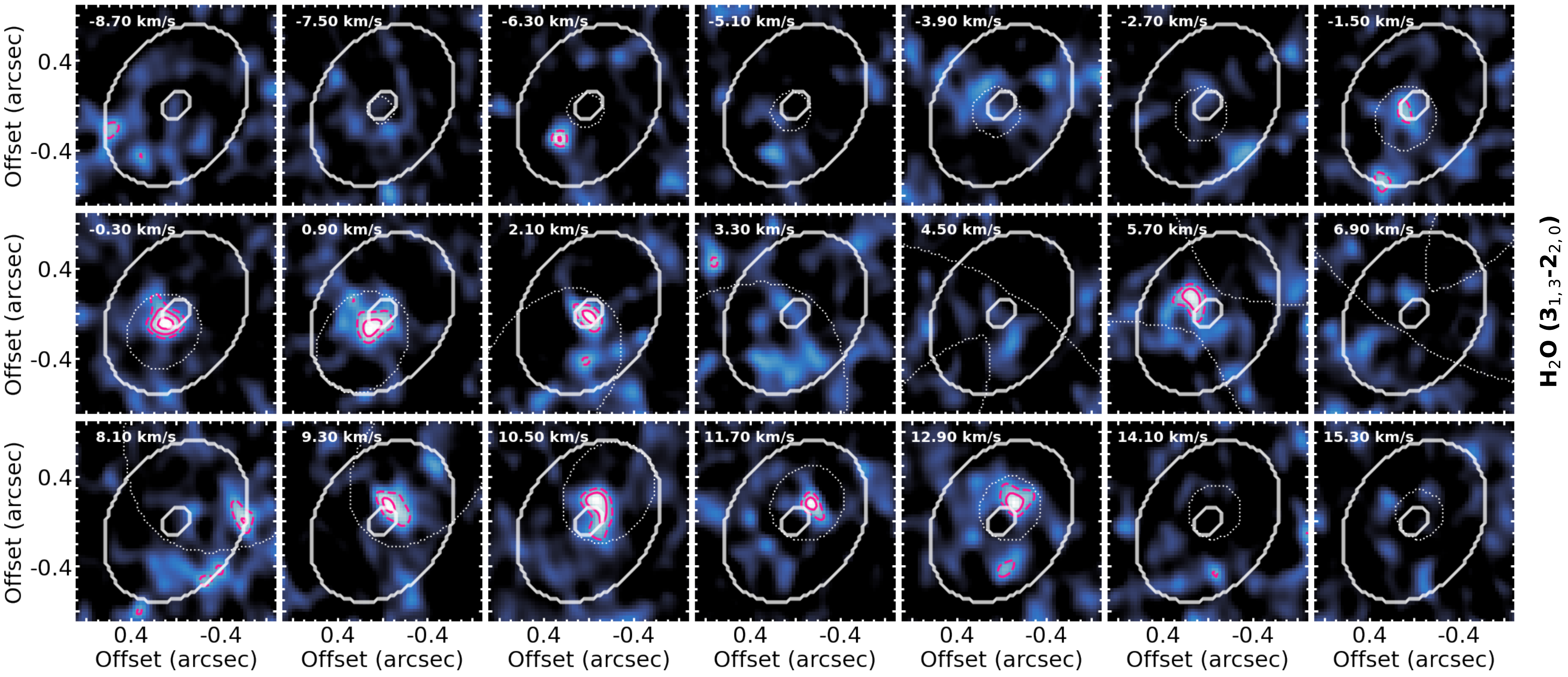}
    \caption{Channel maps of the water line at 183~GHz. The systemic velocity is 5.7~\kms. White ellipses correspond to the location of the two peaks in the radial profile (see Fig.~\ref{fig:radial_profile}) at $R = 15$ and $85$~au. Pink contours show 3,4, and 5$\sigma$ levels, where $\sigma$ is evaluated as standard deviation inside an annulus between 250 and 450 au for each channel. The white dashed contours show the expected Keplerian pattern.}
    \label{fig:channel_maps}
\end{figure*}

Figure~\ref{fig:channel_maps} shows the channel maps of the detected water line in the HD~100546 disk. The cube was imaged with natural weighting and a channel width of $1.2$~\kms (see Sect.~\ref{subsec:imaging} for more details on the imaging procedure). 

\section{Water observations with \textit{Herschel} and ALMA} \label{app:model_ncr}

\begin{figure*}[ht!]
    \centering
    \includegraphics[scale=0.25]{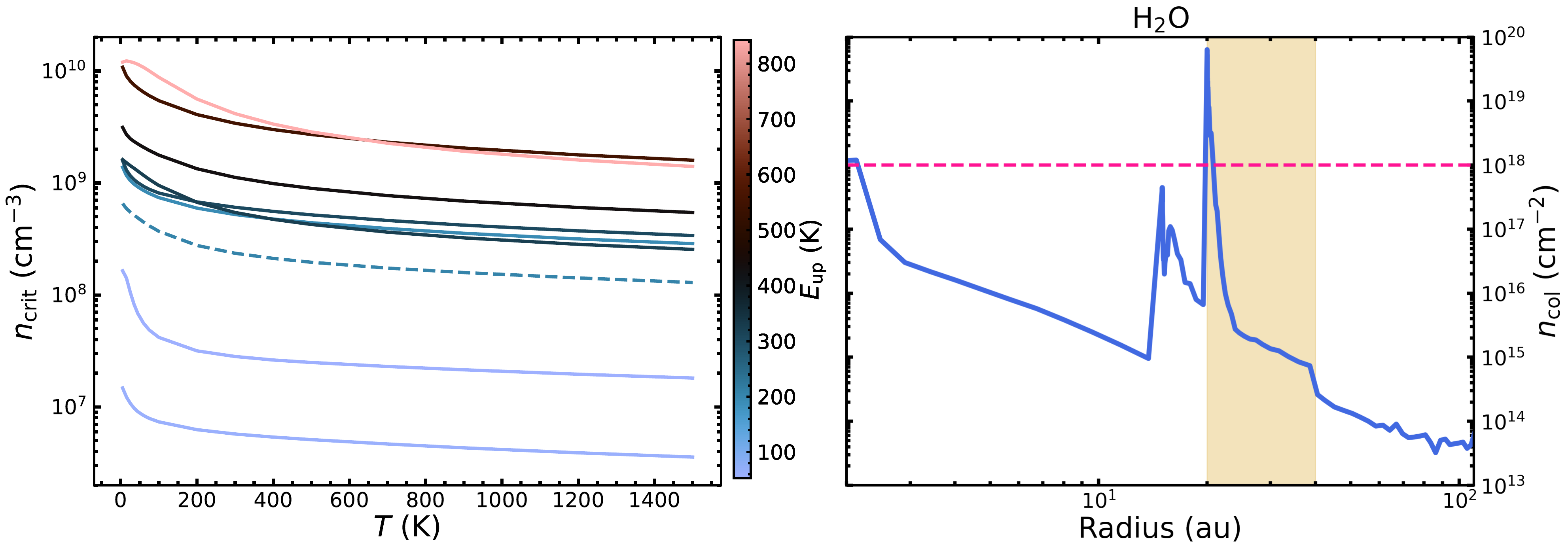}
    \caption{Left: critical density as a function of temperature for the \textit{Herschel} lines included in \cite{pirovano2022} (solid lines) and the ALMA 183~GHz line (dashed). The colorbar shows the upper state energies of each transition. Right: Water column density from the DALI model of the HD~100546 disk from \cite{leemker2024}. The pink horizontal line indicates the density above which water self-shielding is expected to be important. The brown shaded region indicate the dust ring between 20 and 40~au.}
    \label{fig:model_water_ncr}
\end{figure*}

The left panel of Fig.~\ref{fig:model_water_ncr} shows the critical density of the Herschel lines in \cite{pirovano2022} (solid lines) and of the ALMA 183~GHz line (dashed line), as a function of temperature. The profiles are colored based on their upper state energy. The critical densities are calculated following \cite{faure2024}:
\begin{equation}
    n_{\mathrm{crit}} = \frac{\sum_{k} A_{jk}}{\sum_{k} \kappa_{jk}} = \frac{\sum_{k} A_{jk}}{\sum_{k < j} \gamma_{jk} + \sum_{k > j} \frac{g_k}{g_j} \gamma_{kj} e^{-\left(E_k - E_j\right)/kT}},
\end{equation}
where $A_{jk}$ is the Einstein coefficient of the spontaneous emission and $\gamma_{jk}$ is the collision rate out of the level $j$ into the level $k$. The upward collision rate $\gamma_{jk}$ is related to the downward $\gamma_{kj}$ one as follow \citep{shirley2015}:
\begin{equation}
    \gamma_{jk} = \frac{g_k}{g_j} \gamma_{kj} e^{-\left(E_k - E_j\right)/kT},
\end{equation}
where $g$ is the degeneracy, $E$ is the energy, and $T$ is the kinetic temperature. The critical density is calculated under the assumption of optically thin water emission, and neglecting absorption, stimulated emission, and background emission.

The right panel of Fig.~\ref{fig:model_water_ncr} shows the radial profile of the water column density obtained from the thermochemical model built by \cite{leemker2024} using DALI \citep{bruderer2012dali1,bruderer2013dali2} for the HD~100546 disk. We highlight that this model aimed at reproducing the emission morphology of a variety of molecules detected with ALMA, in particular CO isotopologues, HCN, CN, \ce{C2H}, NO, and \ce{HCO^+}, but it was not specifically built to reproduce the water emission and observations in this source.

\bibliography{sample7}{}
\bibliographystyle{aasjournalv7}

%% This command is needed to show the entire author+affiliation list when
%% the collaboration and author truncation commands are used.  It has to
%% go at the end of the manuscript.
%\allauthors

%% Include this line if you are using the \added, \replaced, \deleted
%% commands to see a summary list of all changes at the end of the article.
%\listofchanges

\end{document}